\documentclass[11pt]{article}

\begin{document}

\title{On the hyperfine anomaly in Eu isotopes}
\author{J. R. Persson \\
Department of Natural Sciences\\
University of Agder\\
Servicebox 422\\
N-4604 Kristiansand\\
Norway\\
jonas.persson@uia.no}
\maketitle

\begin{abstract}
A new method for determining the hyperfine anomaly,
without knowing the nuclear magnetic moment, is used for the first
time on a series of unstable isotopes. The relative large number of
experimental data in Eu makes it possible to determine the hyperfine
anomaly for a number of unstable isotopes. Calculations of the Bohr-Weisskopf
effect and hence the hyperfine anomaly has been performed using the
particle-rotor formalism. The result from the calculations and experiments is compared
with other theoretical calculations and the empirical Moskowitz-Lombardi
formula.

\end{abstract}

%\PACS{{31.30 Gs, 32.10 Fn}  {}

\section{Introduction}

The study of hyperfine structure (hfs) in atoms has provided information of
the electromagnetic moments of the nucleus as well as information on electron
properties' \cite{otten,lindgrenrosen}. The magnetic hfs has in addition
proved to give information on the distribution of magnetisation in the nucleus
through the so called Bohr-Weisskopf effect (BW-effect)
\cite{bohrweisskopf,Fujitaarima,Buttgenbach}. The influence of the finite size
of the nucleus on the hyperfine structure was first considered by Bohr and
Weisskopf \cite{bohrweisskopf}. They calculated the hyperfine interaction
(hfi) of s$_{1/2}$ and p$_{1/2}$ electrons in the field of an extended
nucleus, and showed that the magnetic dipole hyperfine interaction constant
($A$) for an extended nucleus is generally smaller than that expected for a
point nucleus. The extended charge distribution of the nucleus
gives rise to the so-called Breit-Rosenthal effect
(BR-effect)\cite{Breit,Crawford,Pallas,Rosenberg}. In this case, as in most cases, the
differential BR-effect is negligible when two isotopes are compared. Inclusion
of the BR-effect will not have any influence on the results, since the BW- and
BR-effects show the same behaviour. The BR-effect is therefore neglected in
the following discussion. Isotopic variations of magnetic moments became larger than
those in the point dipole interaction when there are different contributions
to the hfs from the orbital and spin parts of the magnetisation in the case of
extended nuclei. The fractional difference between the point nucleus hfi
constant ($A_{point}$) and the constant obtained for the extended nuclear
magnetisation is commonly referred to as the Bohr-Weisskopf (BW) effect
\cite{Buttgenbach}.The hfi constant $A$ can therefore be written as%

\begin{equation}
A=A_{point}\left(  1+\epsilon_{BW}\right) \label{eq1}%
\end{equation}

where $\epsilon_{BW}$ is the BW-effect, and $A_{point}$ is the $A$ constant for
a point nucleus. The BW-effect is dependent on both nuclear properties as well
as atomic properties, i.e. the electron density within the nucleus. The
nuclear part, i.e. the distribution of nuclear magnetisation, can be
calculated using different nuclear models \cite{Fujitaarima,Buttgenbach}.
Because electronic wavefunctions cannot be calculated with high accuracy in
complex atoms, as they can be in hydrogen-like ions and muonic atoms, it is
not always possible to determine\ $\epsilon_{BW}$ directly. However, it is
possible to determine the difference of the BW-effect in two isotopes, the
so-called (differential) hyperfine anomaly (hfa). Where one compares the ratio
of the measured hfs constants for two isotopes with the independently measured
ratio of the nuclear magnetic dipole moments to extract the hfa,$^{1}%
\Delta^{2}$, for the isotopes 1 and 2, and a given atomic state:%

\begin{equation}
1+{^{1}\Delta^{2}}={{\frac{A^{(1)}}{A^{(2)}}}{\frac{{\mu_{I}^{(2)}/I^{(2)}}%
}{{\mu_{I}^{(1)}/I^{(1)}}}}\approx1+{\epsilon_{BW}^{(1)}}-{\epsilon_{BW}%
^{(2)}}}\label{eq2}%
\end{equation}

where $\mu_{I}$ is the nuclear magnetic dipole moment, and I the nuclear spin.
For electrons with a total angular momentum j$>$1/2 the anomalies may be disregarded as the corresponding wavefunctions vanish
at the nucleus. The hfa can show a dependence of the atomic state, a state
dependent hfa, where the values for different states can vary significantly.
The reason is that the hyperfine interaction consists of three parts
\cite{Ref4,Ref5}, orbital, spin-orbit and contact (spin) interaction, where
only the contact interaction contributes to the hfa. It is suitable to rewrite
the dipole hyperfine interaction constant as%

\begin{equation}
A=A_{nc}+A_{c}\label{eq3}%
\end{equation}

where $A_{c}$ is the contribution due to the contact interaction of s (and
p$_{1/2}$) electrons and $A_{nc}$ is the contribution due to non-contact
interactions. The experimental hfa, which is defined with the total magnetic
dipole hyperfine constant $A$, should then be rewritten to obtain the relative
contact contribution to the hfa:%

\begin{equation}
{^{1}\Delta_{exp}^{2}}={^{1}\Delta_{c}^{2}}{\frac{A_{c}}{A}}\label{eq4}%
\end{equation}

where ${^{1}\Delta_{c}^{2}}$ is the hfa due to the contact interaction, that
is, for an s- or p$_{1/2}$-electron. So far we have considered direct
interactions between the electron and the nucleus, but we should also include
electron-electron interactions. One interaction, which can influence the
hyperfine interaction, is the polarisation of the electron core \cite{Ref4},
which may give a substantial contribution to the experimental hfa
\cite{Buttgenbach}. Core polarisation can be seen as an excitation of a
d-electron, which will not itself give any contribution to the hfa, to an s-electron,
which gives a large hfa. Since ${^{1}\Delta_{s}^{2}}$, is independent of n in the first approximation, it
is possible to use it to obtain values of the core-polarisation
\cite{Buttgenbach,persson1}.

From the discussion, one is led to the conclusion that one needs independent
measurements of the nuclear magnetic moments and the $A$-constants in order to
obtain the hfa, however, this is not true. It has been shown by Persson
\cite{persson} that it is possible to extract the anomaly solely from the
$A$-constants of two different atomic levels, provided the ratio $\left(  \frac{A_{s}}{A}\right)
$differs substantially for the different levels. Comparing the A-constants
ratio, for two isotopes, in two atomic levels, gives:%

\begin{equation}
{\frac{{A_{B}^{(1)}/A_{B}^{(2)}}}{{A_{C}^{(1)}/A_{C}^{(2)}}}}\approx
{1+{^{1}\Delta_{s}^{2}}({{\frac{A_{s}^{B}}{A^{B}}}-{\frac{A_{s}^{C}}{{A^{C}}}%
})}}\label{eq5}%
\end{equation}

Where B and C denotes different atomic levels and 1 and 2 denotes different
isotopes. The ratio between the two $A$-constant ratios for the isotopes will
only depend on the difference of the contact contributions of the two atomic
levels and the hfa for the s electron. It should be noted that the ratio
$\left(  \frac{A_{s}}{A}\right)  $ is isotope independent. Once determined for
one isotopic pair, the ratio can be used for all pairs, which is useful in the
study of hfa in radioactive isotopes. The ratio can be determined in two
different ways; either by making an analysis of the hyperfine interaction or
by using a known hfa as a calibration. It should be pointed out that the
atomic states used must differ significantly in the ratio$\left(  \frac{A_{s}%
}{A}\right)  $, as a small difference will lead to an increased sensitivity to
errors, as can be deduced from eqn \ref{eq5}.

Since the hfa is normally very small (1\% or less) it is
necessary to have high accuracy, better than 10$^{-4}$ \cite{Buttgenbach}. In
the case of stable isotopes there is no major problem to measure the nuclear
magnetic moment, with NMR or ABMR, while unstable isotopes are more difficult.
In most cases there does not exist any high precision measurements of the
nuclear magnetic moment. However, there might exist measurements of two
$A$-constants, if the unstable isotopes nuclear charge radius has been
measured by means of laser spectroscopy \cite{otten}. In order to obtain the
hfa one needs to measure the $A$-constants with an accuracy better than
10$^{-4}$. This can be done by laser spectroscopy when the $A$-constant is
larger than about 1000 MHz.
 
As pointed out in a review on hfa \cite{Buttgenbach}, off-diagonal hyperfine interactions may simulate a hfa. However, corrections due to off-diagonal hyperfine interaction affects mainly the electric quadrupole hyperfine interaction constants, unless the correction is very large, in which case the experimental error in the $A$-constants is large. In most cases is the correction smaller than the experimental error and can be neglected, especially with laserspectroscopy where the error is on the order of 1 MHz. In the present study are the experimental error in the $A$-constants so large that off-diagonal hyperfine interaction corrections can be neglected.

\subsection{Hyperfine structure measurements in Eu}

Europium has been subject to many investigations since the measurements of
Sch\"{u}ler and Schmidt in 1935 \cite{schuler}. Since then have a lot of
measurements been performed using a variety of different methods, for example;
Fabry-Perot spectroscopy \cite{muller}, atomic beam magnetic resonance
\cite{evans}, Level crossing spectroscopy \cite{champeau}, laser atomic beam
spectroscopy\cite{zaal,eliel}, laser-rf double resonance \cite{sen} and in an
ion trap \cite{becker}. In total has the hfs been determined in over 30 atomic
states and 15 states in $Eu^{+}$. The high accuracy in some measurements has
made it possible to determine the hyperfine anomaly. One problem with a
complex atom like Eu is that the hyperfine anomaly is state dependent and has
to be analysed to give the correct value \cite{Buttgenbach}. A case study of
the hyperfine anomaly was done by B\"{u}ttgenbach in his review
article \cite{Buttgenbach}, where he found the values of the s-electron
hyperfine anomaly ($^{151}\Delta_{s}^{153}$)to be -0.64(3)\%, -0.66(3)\% and
-0.59(5)\% depending on the states and experimental method used. The average
value -0.64(4)\% is probably a very good approximation, since the value
obtained from Fabry-Perot measurements in the ion (-0.59(5)\%) is probably too
low. The accurate measurements using an ion trap by Becker et al.
\cite{becker} obtained an uncorrected, i.e. state dependent, value of
-0.663(18)\% for the hyperfine anomaly in the ground state of $Eu^{+}$. The
uncertainty in this value originates mainly from the magnetic moment
measurements \cite{evans}. There is no doubt that the high accuracy of the ion
trap will give very precise values for the hyperfine anomaly, once the
accuracy of the nuclear magnetic dipole moment can match the accuracy of the
$A$-constant ($\approx10^{-8}$) in ion trap measurements\cite{Trapp} or if the
$A$-constant can be measured in another level, atomic or double-ionised, with
matching accuracy.

In a paper by Asaga et al.\cite{asaga} a theoretical study of the hyperfine
anomaly in odd isotopes of Eu was performed. They addressed the question of
the universality of the empirical Moskowitz-Lombardi formula \cite{ML}. In
this article the method of Persson \cite{persson} is applied to the
measurements of Ahmad et al. \cite{ahmad} and H\"{u}hnermann et al.
\cite{huhnermann} in order to give an estimate on the validity of the
Moskowitz-Lombardi formula in Eu. In addition a calculation of the hyperfine
anomaly using the particle-rotor model is presented

\section{Hyperfine anomaly in unstable Eu isotopes}

The unstable Eu isotopes have been studied using laser spectroscopy by Ahmad
et al.\cite{ahmad}, H\"{u}hnermann et al.\cite{huhnermann} and by D\"{o}rschel
et al. \cite{dorschel}. Measurements has also been performed by Enders et al
\cite{enders,enders1,enders2} using a Paul-trap, to obtain the hyperfine
structure constants in the ground and first metastable states of $Eu^{+}$. The
high precision values from the ion trap measurements can not, for the time
being, be used to extract the hyperfine anomaly as the s-electron contribution
is almost the same in these levels \cite{persson}.

\subsection{\bigskip Measurements done by H\"{u}hnermann et al.}

H\"{u}hnermann et al.\cite{huhnermann} obtained information of the hyperfine
anomaly by defining an angular factor $f(1,2)$, where 1 and 2 are two
different isotopes, which was found from a fit of the experimental
$A$-constants. The factor $f(1,2)$ was defined as:%

\begin{equation}
{{^{1}\Delta_{\exp}^{2}=}-c_{s}\cdot f(1,2)/a^{(2)}}%
\end{equation}

where $c_{s}={\frac{{(J(J+1)-L(L+1)+S(S+1)}}{{2J(J+1)}}}$ and ${a^{(2)}}$ the
experimental $A$-constant for a state in isotope 2. In this way an experimental value
$f(151,153)$ = -0.551(3) MHz was obtained \cite{huhnermann}.

The factor $f(1,2)$ can be expressed using the normal nomenclature, with the
experimental, i.e. state dependent, hyperfine anomaly $^{1}\Delta_{exp}^{2}$,
as\cite{persson},
\begin{equation}
{^{1}\Delta_{exp}^{2}={{\frac{a_{s}}{a_{exp}}}}\cdot{^{1}\Delta_{s}^{2}}}%
\end{equation}
and the factor $c_{s}$ as the total angular part of the contact interaction in
the parametrisation of the hfs \cite{lindgrenrosen}, provided that the states
investigated are purely LS-coupled. This leads to
\begin{equation}
{f(1,2)=a_{s}^{(2)}\cdot^{1}\Delta_{s}^{2}.}%
\end{equation}
Note that the factor ${a_{s}^{(2)}}$ must be corrected, so that it only takes
into account the angular parts for the electrons influenced by the hyperfine
anomaly, that is s-electrons. The states H\"{u}hnermann et al. studied, the
$4f^{7}5d\>^{9}\!D_{J=2...9}$ in $Eu^{+}$, are close to pure LS-coupling.
However, there are no free s-electrons in this configuration why the hyperfine
anomaly arises from core-polarisation. The core polarisation can be expressed
as an excitation of an d-electron to an s-electron, thus showing a hyperfine
anomaly. Since the hyperfine anomaly, due to core-polarisation, show the same
angular dependence as the contact interaction, one can use the hyperfine
anomaly to find the core-polarisation \cite{Buttgenbach,persson1}. The factor
$a_{s}^{(2)}$ should then be the core-polarisation contribution to the hfs. An
analysis of this case has been performed by Persson \cite{persson1}. Using his
value ( $a_{s}^{(2)}=-669MHz$) and the angular part for the d-electron
($\frac{1}{8}$ of the total angular contribution), one finds that%

\begin{equation}
{f(1,2)=-669/8\cdot}^{1}{\Delta_{s}^{2}.}%
\end{equation}

yielding ${{^{151}\Delta_{s}^{153}=-0.659(4)\%}}$, in agreement with earlier
results. The values of $f(1,2)$ for other isotopes from H\"{u}hnermann et
al.\cite{huhnermann} are given in table \ref{tab:1} together with the derived
hyperfine anomaly, $^{1}\Delta_{s}^{2}$. Note that the sign of $f(1,2)$ is
different in the original article, since the factor $c_{s}$ is defined in a
different way in this article.

\begin{table}[ptbh]
\caption{Hyperfine anomalies for Eu isotopes, obtained from the work of
H\"{u}hnerman et al. \cite{huhnermann} }%
\label{tab:1}%
\begin{tabular}
[c]{lll}\hline
A & $f(151,A)$ & ${^{151}\Delta_{s}^{A}}(\%)$\\\hline
145 & 0.07(20) & -0.08(24)\\
146 & -0.11(37) & 0.13(44)\\
147 & -0.31(40) & 0.37(48)\\
151 & 0 & 0.00\\
152 & -0.42(5) & 0.50(6)\\
153 & 0.551(3) & -0.659(4)\\\hline
\end{tabular}
\end{table}

\subsection{\bigskip Measurements done by Ahmad et al.}

The values of Ahmad et al. \cite{ahmad} are used directly together with
equation (3) in order to obtain the hyperfine anomaly. In their study of
nuclear spins, moments and changes of the mean square charge radii, they used
two atomic transitions, 459.4 nm and 462.7 nm, connecting the atomic ground
state $(4f^{7}6s^{2}\;{^{8}}\!S_{7/2})$ with two excited states $(4f^{7}%
6s6p\;$ $^{8}\!P_{9/2,7/2})$. Since these transitions have been studied with higher
accuracy by Zaal et al. \cite{zaal}, there exists a calibration of the hyperfine
anomaly for the stable isotopes.

The hfs of the ground state was not resolved in the measurements \cite{ahmad}
why the $A$-constants of the excited states are used. Using the measured
$A$-constants for the $^{8}P_{9/2,7/2}$ states from \cite{ahmad} and eqn.(3)
it is possible to deduce the hyperfine anomaly, for most unstable isotopes,
using the -0.64(4)\% value for the hyperfine anomaly between the stable
isotopes as a calibration. The result is presented in table \ref{tab:2}.
Comparing the results in table \ref{tab:1} and \ref{tab:2}, shows an agreement
within errors. The values for $^{147}Eu$, still within errors, differs in
sign, something that can be explained by the large experimental errors.

\begin{table}[ptbh]
\caption{Magnetic moments and hyperfine anomalies for Eu isotopes, obtained
from Ahmad et al.\cite{ahmad} }%
\label{tab:2}%
\begin{tabular}
[c]{llll}\hline
A & I & $\mu_{I}$ & ${^{151}\Delta_{s}^{A}}(\%)$\\\hline
142 & 1 & 1.536(19) & -0.14(1.18)\\
142m & 8 & 2.978(11) & -0.08(31)\\
143 & 5/2 & 3.673(8) & -0.06(17)\\
144 & 1 & 1.893(13) & -0.19(48)\\
145 & 5/2 & 3.993(7) & -0.08(15)\\
146 & 4 & 1.425(11) & 0.12(50)\\
147 & 5/2 & 3.724(8) & -0.12(17)\\
148 & 5 & 2.340(10) & 0.08(31)\\
149 & 5/2 & 3.565(6) & -0.19(16)\\
150 & 5 & 2.708(11) & 0.08(28)\\
151 & 5/2 & 3.4717(6) & 0.00\\
153 & 5/2 & 1.5330(8) & -0.64(4)\\\hline
\end{tabular}
\end{table}

\bigskip

As can be seen the errors are larger than the actual values except for
$^{149}Eu$ (table 2) and $^{152}Eu$ (table 1) which makes it difficult to draw
any deeper conclusions, other than the general trends.

From the experimental values in table 2, there seems to be an odd-even
staggering which changes sign at the N=82 (A=145) neutron shell closure, but
the uncertainties are too large to be able to be sure about this. The drastic
change in nuclear magnetic dipole moment between N=88 (A=151) and N=90 (A=153)
due to the shape transition is also reflected in the hyperfine anomaly. The
hfa for the lighter isotopes is fairly constant, indicating that the
magnetisation does not change much from the spherical $^{145}$Eu nucleus to
$^{151}$Eu. With these experimental results we can make comparisons with
theoretical calculations.\bigskip

\section{Calculations of the hyperfine anomaly}

The Bohr-Weisskopf effect, and thereby the hyperfine anomaly, was investigated
by making particle-rotor calculations based on the modified oscillator
(Nilsson) potential, using standard parameters \cite{semmes} as far as
possible. As the nuclear magnetic moment and the Bohr-Weisskopf effect
calculations are mainly analogous, it is sensible to adjust the parameters in
the calculation so that both the energy levels and nuclear magnetic moments
fits well with experimental values. The calculated hyperfine anomalies in the
odd isotopes are given in table 3. As expected the Bohr-Weisskopf effect, and
thus the hyperfine anomaly, stays fairly constant from A=145 to A=151 with an
abrupt change at the shape transition between A=151 to 153.

For the odd isotopes Asaga et al.\cite{asaga} have done a theoretical study
and calculated the BW-effect, and thus the hyperfine anomaly. Since they
addressed the question on the validity of the empirical Moskowitz-Lombardi
formula \cite{ML}:%

\begin{equation}
{{\epsilon_{BW}}= {\frac{\alpha}{\mu_{I}}}},
\end{equation}

has the hyperfine anomaly \ also been calculated using this formula. The
constant $\alpha$ has been taken to be 0.015 n.m. It should be noted that this
value of $\alpha$ is close to the value of Hg \cite{ML} and close to the
values obtained for Ir \cite{Butt,moz}and Au\cite{ekstrom}, however, the sign
is different.

The results are presented in table 3 together with the experimental values
obtained from the measurements of Ahmad et al.\cite{ahmad}.
\begin{table}[h]
\caption{Hyperfine anomaly in Eu, experimental and calculations. MO denotes
values obtained from Particle-rotor calculations, I and II values from
\cite{asaga} and ML values obtained from the Moskowitz-Lombardi formula
\cite{ML} }%
\label{tab:7}
\begin{center}%
\begin{tabular}
[c]{llllll}\hline
& $^{151}\Delta_{\exp}^{A}(\%)$ & $^{151}\Delta_{MO}^{A}(\%)$ & $^{151}%
\Delta_{I}^{A}(\%)$ & $^{151}\Delta_{II}^{A}(\%)$ & $^{151}\Delta_{ML}%
^{A}(\%)$\\\hline
& Exp. & Calc. & Calc. & Calc. & Calc.\\\hline
$^{145}\!Eu$ & -0.08(15) & 0.000 & 0.021 & 0.031 & 0.056\\
$^{147}\!Eu$ & -0.12(17) & 0.002 & 0.010 & 0.008 & 0.029\\
$^{149}\!Eu$ & -0.19(16) & 0.002 & 0.007 & 0.006 & 0.011\\
$^{151}\!Eu$ & 0 & 0.0 & 0 & 0 & 0.000\\
$^{153}\!Eu$ & -0.64(4) & -0.768 & -0.127 & 0.003 & -0.546\\
$^{155}\!Eu$ &          & -0.768 & -0.127 & -0.001 & -0.555\\\hline
\end{tabular}
\end{center}
\end{table}

Clearly the theoretical calculations manage to reproduce the trend of the
hyperfine anomaly, even if the values from Asaga et al.\cite{asaga} are too
small. It is also interesting to note that the empirical M-L formula is still
valid for Eu and not only for elements around Z=80 (Ir,Au,Hg). However the
change in sign of $\alpha$ is not explained.

From the operators describing the nuclear magnetic moment and the
Bohr-Weisskopf effect, one can easily see that the M-L formula is justified to
some extent. However the empirical fit inplies a state independence of the
spin operator which is not in accordance with theoretical predictions. The
question of the universal validity of the empirical Moskowitz-Lombardi formula
is therefore still open.

\bigskip

\begin{table}[ptb]
\caption{Magnetic moments and Bohr-Weisskopf effect for odd Eu isotopes. MO
denotes values obtained from particle-rotor calculations, I and II values from
\cite{asaga}.}%
\begin{tabular}
[c]{llllll}\hline
A & $\mu_{I}\left(  exp\right)  $ & $\mu_{I}\left(  MO\right)  $ & ${\epsilon
}(\%)\left(  MO\right)  $ & $\epsilon{_{.}}(\%)$I & $\epsilon_{.}%
(\%)$II\\\hline
145 & 3.993 & 3.773 & -1.001 & -1.067 & -1.067\\
147 & 3.724 & 3.720 & -1.003 & -1.056 & -1.044\\
149 & 3.565 & 3.552 & -1.003 & -1.053 & -1.042\\
151 & 3.4717 & 3.506 & -1.004 & -1.046 & -1.036\\
153 & 1.5330 & 1.532 & -0.236 & -0.919 & -1.039\\
155 & 1.52 & 1.529 & -0.236 & -0.919 & -1.035\\\hline
\end{tabular}
\end{table}

If we inspect the actual values of the Bohr-Weisskopf effect instead of just
looking at the hyperfine anomaly, we find a significant difference in the
values obtained from the two methods of calculation. As we have an
experimental value of the BW-effect in $^{151}$Eu from muonic X-ray
measurements, $\epsilon_{BW}=-0.63\left(  13\right)  \%$, it is possible to
further discuss the methods. It is clear that the particle-rotor calculations
show a better agreement with experiment at least for the hyperfine anomaly.
The BW-effect is more difficult to calculate as there seem to be an "offset"
in the calculated results. The agreement with experiment in Eu might be a
coincidence, as preliminary calculations in Gd and Sm show a more complex
situation.\bigskip

\bigskip

\subsection{\bigskip The empirical Moskowitz-Lombardi formula.}

The empirical ML formula was established in 1973 as a rule for the s-electron
BW-effect in mercury isotopes\cite{ML}.%

\begin{equation}
\epsilon_{BW}{=}\frac{{\alpha}}{\mu_{I}}, \alpha= \pm 1.13\cdot10^{-2}\mu_{N}, I=l\pm\frac{1}{2}%
\end{equation}

where $l$ is the orbital momentum for the odd neutron. It turned out that the
empirical rule provided a better agreement with experimental hfa than the
theoretical calculations performed by Fujita and Arima \cite{Fujitaarima}
using microscopic theory. The rule can be qualitatively explained by the
microscopic theory used by Fujita and Arima \cite{Fujitaarima}, where the
parameter $\alpha$ is more state independent than given by the theory. Further
investigations gave an analogous expression for the odd-proton nuclei
$^{191,193}$Ir , $^{197,199}$Au and $^{203,205}$Tl, but also for the
doubly-odd $^{196,198}$Au nuclei. The results indicate that the spin operators
$g_{s}^{\left(  i\right)  }\Sigma_{i}^{\left(  1\right)  }$are state
independent for these nuclei. It is worth noting that all nuclei discussed lie
close to the doubly closed shell nucleus $^{208}$Pb, where one would expect
the single particle model to provide a good description of the nucleus. It is
not apparent that the rule is applicable to lighter nuclei.

With the data presented here it is possible to make a comparison with
lanthanide nuclei. As has been shown in Eu, the ML formula seems to account
for the hfa, even though the obtained value of the BW-effect differs more from
the experimental value of $^{151}$Eu. It should be noted that the sign of
$\alpha$ is different from the value obtained for nuclei close to $^{208}$Pb,
indicating that the ML rule is not universal. In order to further test the ML
formula, the values of $\alpha$ was deduced from the experimental values of
the hfa and nuclear magnetic moments in Nd, Gd \cite{persson2} and Eu, using:%

\begin{equation}
^{1}\Delta^{2}=\alpha\left(  \frac{1}{\mu_{I,1}}-\frac{1}{\mu_{I,2}}\right)
\end{equation}

If the ML rule show some sort of general validity the values of $\alpha$
should stay fairly constant and show a different sign between Eu (odd-proton)
and Nd and Gd (odd-neutron). The obtained values are shown in table 5. As can
be seen there are no indications that the ML rule is applicable for these
nuclei. The conclusion would be that one should be very careful in applying
the ML rule for lighter nuclei.

\begin{table}[h]
\caption{Hyperfine anomaly in the lanthanides.}%
\label{tab:6}
\begin{center}%
\begin{tabular}
[c]{lcccc}\hline
& $\mu_{I,1}$ & $\mu_{I,2}$ & ${^{1}\Delta_{s}^{2}}\left(  \%\right)  $ &
$\alpha\left(  10^{-2}\right)  $\\\hline
$_{60}^{143,145}Nd$ & -1.065 & -0.656 & 0.2034 & $0.35$\\
$\!_{63}^{151,153}Eu$ & 3.4717 & 1.533 & -0.64 & $1.76$\\
$\!_{64}^{157,155}Gd$ & -0.3387 & -0.2572 & 0.106 & $0.11$\\\hline
\end{tabular}
\end{center}
\end{table}

\section{Conclusions}

The new method of Persson\cite{persson} has been applied to Eu, and provided
preliminary values of the hyperfine anomaly, the experimental data are not precise enough for all isotopes. The values obtained are in
agreement with the particle-rotor calculations and the theoretical predictions
of the trends by Asaga et al. \cite{asaga}. A comparison with the empirical
M-L formula shows that it can be used for Eu, however the constant $\alpha$
attains a different sign compared with the values for Ir, Au and Hg. There
also seems to exist an odd-even staggering of the hyperfine anomaly in Eu,
similar to what was found in Au \cite{ekstrom}. This analysis shows that there
is a need for further studies of the hyperfine anomaly in Eu. The application
of ion traps in measuring the $A$-constants of unstable Eu isotopes
\cite{enders1,enders2} has shown an excellent accuracy and will, when high
accuracy measurements of the nuclear magnetic dipole moments are available or
$A$-constants in suitable atomic or ionic levels, give a deeper understanding
of the hyperfine anomaly in Eu and hopefully to all nuclei.

As has been shown in the case of Eu, it is possible to obtain a lot of
information on the hfa without knowing the nuclear magnetic moment of the
isotopes under study. It has also been shown that the ML rule is not
universal, why one has to be careful in applying it to nuclei far from the
doubly closed shell nucleus $^{208}$Pb.

\end{document}